\documentclass[pra,aps,reprint,showpacs,superscriptaddress,amsmath]{revtex4-1}

\usepackage{bm}
\usepackage{graphicx}
\usepackage{dsfont}
\usepackage[usenames,dvipsnames]{xcolor}

\newcommand{\bra}[1]{\langle #1 | \,}
\newcommand{\ket}[1]{\, | #1 \rangle}
\newcommand{\braket}[2]{\langle #1 | #2 \rangle}
\newcommand{\expv}[1]{\langle #1 \rangle}
\newcommand{\om}{\omega}
\newcommand{\Om}{\Omega}
\newcommand{\Omp}{\Omega^\prime}
\newcommand{\ga}{\gamma}

\newcommand{\de}{\delta}
\newcommand{\De}{\Delta}

\newcommand{\sgeh}{\hat{\sigma}_{ge}}

\newcommand{\sgsh}{\hat{\sigma}_{gs}}
\newcommand{\sesh}{\hat{\sigma}_{es}}
\newcommand{\ssep}{\hat{\sigma}_{se^\prime}}

\newcommand{\segh}{\hat{\sigma}_{eg}}
\newcommand{\sepg}{\hat{\sigma}_{e^\prime g}}

\newcommand{\seps}{\hat{\sigma}_{e^\prime s}}

\newcommand{\sepep}{\hat{\sigma}_{e^\prime e^\prime}}
\newcommand{\seeh}{\hat{\sigma}_{ee}}
\newcommand{\sssh}{\hat{\sigma}_{ss}}
\newcommand{\as}{\hat{a}_S}
\newcommand{\ai}{\hat{a}_I}

\newcommand{\dude}[1]{\langle #1 \rangle}

\newcommand{\ggs}{\ensuremath{\gamma _{\text{gs}}}}
\newcommand{\gge}{\ensuremath{\gamma _{\text{ge}}}}
\newcommand{\gse}{\ensuremath{\gamma _{\text{se}}}}
\newcommand{\Op}{\ensuremath{\Omega \sp{\prime}}}
\newcommand{\Ops}{\ensuremath{\Omega \sp{\prime *}}}
\newcommand{\D}{\ensuremath{\Delta}}

\newcommand{\sge}{\ensuremath{\sigma _{\text{ge}}}}

\newcommand{\sgs}{\ensuremath{\sigma _{\text{gs}}}}

\begin{document}

\title{Fidelity of photon propagation in 
electromagnetically induced transparency in the presence of four-wave mixing}

\author{Nikolai Lauk}
\email{nlauk@physik.uni-kl.de}

\author{Christopher O'Brien}

\author{Michael Fleischhauer}

\affiliation{Department of Physics and research center OPTIMAS
University of Kaiserslautern, D-67663 Kaiserslautern, Germany}

\date{\today}

\begin{abstract}

We study the effects of the four-wave mixing (4WM) in a quantum memory scheme based on
electromagnetically induced transparency (EIT). We treat the problem of field propagation on the quantum mechanical level, which allows us to calculate the fidelity for propagation for a quantum light pulse such as a single photon. 
While 4WM can be beneficial for classical, all-optical information storage, the quantum noise associated with the signal amplification
and idler generation is in general detrimental for a quantum memory. We identify a range of
parameters where 4WM makes a single photon quantum memory impossible. 

\end{abstract}


\maketitle
\section{Introduction}

A reliable quantum memory for photons is one of the essential ingredients for quantum networks and optical quantum computing. 
There have been several proposals for photon storage, which fall into three main categories: photon echo based techniques \cite{echorev}, 
far detuned Raman systems \cite{ramanmem}, and electromagnetically induced transparency (EIT) \cite{Fleischhauer_2005}. 
In all these schemes the storage of single photons occurs by mapping the quantum state of photons onto a long lived 
atomic excitation.

In this paper we will focus on the EIT based scheme, where one uses a strong control field to couple an incoming signal pulse to the atomic spin coherence resulting in the common propagation of 
both as a dark state polariton. By adiabatically switching off the control field the signal field is mapped on the spin coherence and later, after some storage time, is retrieved by switching on the control field. 
Since its theoretical proposal \cite{Michael_2000, Fleischhauer_2002} and the first experimental realizations \cite{EITexp1,EITexp2} 
there has been a large development of EIT based quantum memories (QM), e.g. successful implementation
in hot gases \cite{hot-gases}, in cold gases using magneto-optical traps (MOT) \cite{cold-gases} or optical lattices \cite{Bloch-storage-exp,lattice}, as well
as solid state systems such as rare-earth doped crystals \cite{doped-crystals}. Using EIT memory, weak coherent pulses have been stored in 
hot Rubidium gas with storage times of $T_s = 1ms$ and storage efficiencies of $45\%$ \cite{Novikova2012}.
While in a cold gas system using the dark MOT technique one could reach storage efficiency of $78\%$ with comparable storage times \cite{coldgas}. 

The storage efficiency of EIT QM is limited by two considerations. First the spatial pulse size $L_p$ must fit entirely inside of the medium $L_p=T_p v_g < L$, 
where $T_p$ is the pulse duration and $v_g$ is the group velocity in the medium, otherwise some of the pulse will leak out during the storage process and be lost.
Secondly, the spectral width of the pulse $\Delta \omega _p$ must be well within the EIT transmission window, $\Delta \omega _p \ll \omega_{EIT} \simeq \sqrt{D} v_g / L$, where $D=L/L_{\rm abs}$ denotes the optical depth of the medium, i.e. the ratio of medium length $L$ to absorption length 
$L_{\rm abs}$ in the absence of EIT.
Since the spectral width and pulse length are inversely proportional $\Delta \omega _p \sim 1/T_p$, 
these two requirements compete with each other and both of them can only be satisfied at large optical depth $\sqrt{D} \gg 1$ \cite{Gorshkov_2007}.

\begin{figure}[t]
\includegraphics[width=6cm]{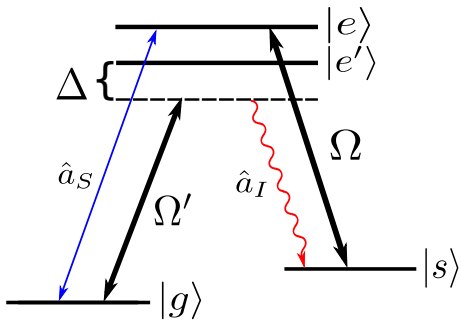}
\caption{
Level scheme for EIT memory with four-wave mixing. A double $\Lambda$-scheme, with one $\Lambda$ being the signal field $\hat{a} _S$ 
and strong control field $\Omega$ giving standard EIT, and a second $\Lambda$ far-detuned from resonance made up of the same control field acting
on the $\ket{g}-\ket{e^\prime}$ transition and the idler field $\hat{a}_I$ generated by 4WM. 
 } 
\label{fig:4levelscheme}
\end{figure}

However with high optical depth, non-linear processes start to become important. 
In particular, a four wave mixing process (4WM) is possible in many of the implementations of EIT QM, where the control field with Rabi frequency $\Omega$ and 
appropriate polarization also acts as a far-detuned field with Rabi frequency $\Op$ on the signal transition spontaneously generating a new `idler' field.
This idler field then moves population into the spin state, which is then pumped by the control field to the excited state 
from where it can come back to the ground state, either by stimulated emission providing amplification to the signal or by spontaneous decay introducing noise, as shown in Fig. \ref{fig:4levelscheme}. 
The medium is still transparent to the signal pulse due to EIT, but now the signal pulse also experiences some gain from 4WM.  

It was originally suggested that 4WM could play a positive role in EIT quantum memories. An experiment \cite{Howell_2009} claimed that 4WM may be useful due to better spatial pulse compression and pulse gain. They also
suggest that with the help of the 4WM one may achieve multimode storage, storing not only the signal mode, but also the idler mode. However this conjecture was disproved in a more
recent experiment \cite{Novikova_2011} where it was clearly shown that multimode storage is not possible in this system, due to no significant slowing of an input idler field, allowing it to escape the medium before storage.
On the other hand, 4WM along with other non-linear effects was used as an explanation of why the storage efficiency has tended to saturate to values lower than 50\% with high $D$ in some EIT QM experiments \cite{Novikova_2008}.

EIT with 4WM has the advantage of signal gain which could be used to compensate losses in the medium, naturally improving the storage of classical signal pulses and thus should not be blamed for the saturation of memory \textit{efficiencies}. 
But the goal of a quantum memory is single photon storage, where gain can become a liability since it is always accompanied by additional noise.
We will show that any benefits of 4WM to EIT QM will be overshadowed by the drawbacks from increased noise generation; noise that will lower the storage \textit{fidelity}. 
We suggest that this additional noise may have already been observed in hot Rb gas experiments such as \cite{Eden1_2009, Eden2_2009}.

Therefore, we address the case of a single photon propagating in an EIT medium
with 4WM, by developing a fully quantum model for pulse propagation in Sect. \ref{sect:model}, which is then solved in Sect. \ref{sect:sol}. 
We then analyze the amplification noise in the system in Sect. \ref{sect:noise} resulting from the spontaneous generation of idler photons coupling
to the signal field.
Our noise analysis is expanded in Sect. \ref{sect:excited} to incorporate noise associated with population decay from
excited states leading to additional fluctuations of atomic dipoles.
In Sect. \ref{sect:results} we use our results to calculate the memory fidelity of EIT with 4WM in
an otherwise loss-less medium. Finally, in Sect. \ref{sect:losses} we consider additional linear losses and
discuss the case where 4WM gain completely compensates these losses in the medium. 

%
%
\section{Model}
\label{sect:model}
We can model the EIT 4WM system as an ensemble of four level atoms, which interact with the strong control field $\vec{E_c}$ coupling the $\ket{s}-\ket{e}$ levels
as well as acting as a far-detuned field on the $\ket{g}-\ket{e'}$
transition and a weak copropagating signal field $\vec{E_S}$ in resonance with the $\ket{g}-\ket{e}$ transition, see Fig. \ref{fig:4levelscheme},
a treatment similar to \cite{Kolchin_2007}. 
Due to the additional coupling of the control field to the $\ket{g}-\ket{e^\prime}$ transition a new idler field will be 
generated which because of the frequency and phase-matching conditions, propagates in the same directions as the other two fields.
In this approach the nature of $\ket{e^\prime}$ depends on the field polarizations. 
If the control and signal fields have the same linear polarization, as is the case in many EIT experiments, then both the signal and $\Omega ^\prime$
couple to the same transition and $\ket{e^\prime}$ may be the same as $\ket{e}$; in this case $\Delta$ is given by the frequency of 
the spin transition $\omega_{s}-\omega_g$.  
Otherwise $\ket{e^\prime}$ is another transition that $\Op$ couples to, and $\Delta$ depends on the frequency of that transition.  
In either case $\ket{e^\prime}$ will eventually be adiabatically eliminated. If there are multiple excited states with large but comparable detunings, $\Delta$ denotes an effective detuning 
resulting from a properly weighted average.
The interaction Hamiltonian in the dipole, rotating wave, and slowly varying envelope approximations is given by:
\begin{align}
&\hat{H}_{\rm int} = \frac{\hbar N}{L}\int_0^L\!\! {\rm d}z\ \Bigl\{\de \seeh +\De \sepep \label{eq:H}\\
& \qquad - \bigl( g_S \as \segh + g_I\ai \seps + \Om \sesh + \Omp \sepg + h.c.\bigr)\Bigr\} \nonumber
\end{align}
where $\De$ is an effective detuning of the control laser from the $\ket{e^\prime}-\ket{g}$ transition, $\de$ is the detuning of the signal photon from the $\ket{e}-\ket{g}$ transition, $\Om=\frac{\mu_{es}E_c}{\hbar}$ 
and $\Omp=\frac{\mu_{e^\prime g}E_c}{\hbar}$ are the Rabi frequencies of the control field with dipole moments $\mu_{eg}$, $\mu_{e^\prime g}$,
$N$ is the number of atoms in the medium, and $L$ is the length of the medium. As in \cite{Fleischhauer_2002}, $\hat\sigma_{\mu\nu}(z)$ are slowly varying continuous atomic ensemble 
spin-flip operators corresponding to the transition from internal state $|\nu\rangle$ to $|\mu\rangle$, and
$\as$ and $\ai$ are dimensionless field operators of the signal- and idler fields, which fulfill
bosonic commutation relations $[\hat a(z),\hat a^\dagger(z^\prime)]= L\delta(z-z^\prime)$.
$g_S=\mu_{eg}\sqrt{\frac{\om_S}{2 \hbar\epsilon_0 V}}$, $g_I=\mu_{e^\prime s}\sqrt{\frac{\om_I}{2 \hbar\epsilon_0 V}}$
are the coupling constants for the field operators $\as$ and $\ai$, where $V$ is the quantization volume.
The equations of motion for the atomic operators are given by the Heisenberg-Langevin equations:
\begin{align}
\frac{\partial}{\partial t}\hat{\sigma}_{ij}=\frac{i}{\hbar}[\hat{H}_{int},\hat{\sigma}_{ij}] - 
\ga_{ij}\hat{\sigma}_{ij} + \delta_{ij}\sum \limits_l r_{li} \hat{\sigma}_{ll}  + \hat{F}_{ij},
\end{align}
where $\ga_{ij}$ are the decoherence rates, $r_{li}$ are the spontaneous emission rates from $\ket{l}$ to $\ket{i}$ and $\hat{F}_{ij}$ are $\de$-correlated Langevin noise operators.
The evolution of the fields is governed by the following propagation equations
\begin{align}
\left(\frac{\partial}{\partial t}+c\frac{\partial}{\partial z}\right)\as&=i g_S N\sgeh,\\
\left(\frac{\partial}{\partial t}+c\frac{\partial}{\partial z}\right)\ai&=i g_I N\ssep.
\end{align}

We simplify these equations by assuming the signal and idler fields remain weak, such that they can be treated perturbatively in the atomic equations. This effectively fixes all of the population in the ground state.
Therefore, the strong control field is not significantly depopulated and we can assume $\Omega$ and $\Op$ are constant. We then adiabatically eliminate $\ket{e'}$ leaving four coupled equations: 
\begin{align}
 &i\partial_t\sgeh = ( \de_{s} \! - \! \de -i\gge )\sgeh - g \as +  i \Om\sgsh  + \! i\hat{F}_{ge}, \label{eq:sge1} \\
 &i\partial_t\sgsh = ( \de_{s} \! - \! \de -i\ggs)\sgsh -  g \frac{\Om^\prime}{\De}\ai^\dagger -  \Om ^* \sgeh  + \! i\hat{F}_{gs}, \label{eq:sgs1} \\
 &\left(\partial_t +c \partial_z\right)\as=igN\sgeh, \label{eq:as1} \\
 &\left(\partial_t +c \partial_z\right)\ai^\dagger=-igN\frac{\Om^{\prime *}}{\De}\sgsh, \label{eq:ai1}
\end{align}
where $\delta _s = |\Op|^2/\Delta$ is the AC-Stark shift and for simplicity we take $g_S = g_I = g$.
The same equations can be derived directly from a 3-level model, as done by Phillips et al. \cite{Novikova_2011}, with the only difference 
being an additional AC-Stark shift in Eq.(\ref{eq:sge1}), $(\delta _s - \delta) \rightarrow (2\delta_s - \delta)$ that can not be removed by the choice of detunings.
Since we are considering the far detuned regime with $\Delta \gg \gge$, this frequency shift $\de_{s}$ can be neglected as it will be much smaller than the EIT transmission window.  
%
\section{Pulse Propagation}
\label{sect:sol}
The EIT QM process consists of a pulse propagating into the medium while the control field is on, 
then adiabatically turning off the control once the pulse is centered in the medium; storing it as a spin excitation.
After some storage time, limited by the spin decoherence rate, the control field is adiabatically switched back on; causing the pulse
to continue to propagate to the end of the medium. Since 4WM only happens when the control field is on, its effects can
be understood by studying the propagating portion of the process. Therefore, we will assume that the control field stays constant, and study
what happens to the signal pulse as it propagates through our medium. Limiting our consideration to propagation however, 
does neglect any losses due to pulse leakage during the storing process. These losses contribute when the compressed pulse length is larger than the
length of the medium.We will show that the pulse compression is to good approximation the same for standard EIT and 4WM EIT. 
Thus the compression losses will be similar for both and we will disregard them here.

In the case of constant control field we can analytically solve Eqs.(\ref{eq:sge1}-\ref{eq:ai1}). For simplicity we take
the single photon detuning to match the AC-Stark shift, $\de = \de_{s} = |\Omega|^2/\Delta$, and set $\ggs = 0$.
The solutions in terms of the optical depth $D$, are given in the frequency domain and co-moving frame as:
\begin{align}
& \hat{a} _S(D,\om) = A(D,\om)\, \hat{a}_S(0,\om) \nonumber \\
& +B(D,\om)\,  \hat{a}_I^\dagger(0,\om) + \delta \hat{\alpha} _S,  \label{eq:as} \\
& \hat{a} _I ^\dagger(D,\om) =  -B(D,\om)\,\hat{a}_S(0,\om) \nonumber \\
& + C(D,\om)\, \hat{a}_I^\dagger(0,\om) + \delta \hat{\alpha} _I. \label{eq:ai}
\end{align}
The coefficients $A(D,\om), C(D,\om)$ describe the spectral transmission for the input signal and idler fields and the coefficient $B(D,\om)$ describes the spectral coupling between the fields.
The corresponding expressions read:
\begin{widetext}
\begin{align}
& A(D,\omega) = \left[ \cosh\Bigl(\frac{D \gge U(\omega)}{2 V(\omega)}\Bigr) + \frac{\gge |\epsilon|^2 -i\omega -i|\epsilon|^2 \omega}{U(\omega)} \sinh\Bigl(\frac{D \gge U(\omega)}{2V(\omega)}\Bigr) \right] 
e^{-\frac{D\gge}{2V(\omega)}\left(i\omega -i\omega |\epsilon|^2 + |\epsilon|^2 \gge\right)}, \label{eq:fullA} \\
& B(D,\omega) =  -\frac{2i\epsilon \Omega}{U(\omega)} \sinh\Bigl(\frac{D\gge U(\omega)}{2V(\omega)}
\Bigr)  e^{-\frac{D\gge}{2V(\omega)}\left(i\omega -i\omega |\epsilon|^2 + |\epsilon|^2 \gge\right)}, \label{eq:fullB} \\
& C(D,\omega) = \left[ \cosh\Bigl(\frac{D \gge U(\omega)}{2 V(\omega)}\Bigr) - \frac{\gge |\epsilon|^2 -i\omega -i|\epsilon|^2 \omega}{U(\omega)} \sinh\Bigl(\frac{D \gge U(\omega)}{2V(\omega)}\Bigr) \right] 
e^{-\frac{D\gge}{2V(\omega)}\left(i\omega -i\omega |\epsilon|^2 + |\epsilon|^2 \gge\right)}, \label{eq:fullC}
\end{align}
\end{widetext}
where $\epsilon = \Op/\Delta$, and:
\begin{align}
& U(\omega) = \sqrt{\bigl[i\omega+|\epsilon|^2(i\omega - \gge) \bigr]^2 + 4|\epsilon|^2|\Omega|^2}, \\
& V(\omega) = \omega (\omega +i\gge) -|\Omega|^2.
\end{align}
The terms $\delta \hat{\alpha} _S$ and $\delta \hat{\alpha} _I$ represent the field fluctuations corresponding to the Langevin noise operators.
These can be neglected in the weak-field approximation since their normal ordered contributions are proportional to $\expv{\seeh}$ and $\expv{\sssh}$ respectively, which are second order in the signal field. 
However since these terms contribute to the noise, they will be considered in Sect. \ref{sect:excited}.

Another important quantity for EIT based QM is the matter excitation, since the light field is mapped onto it during
the storage process. This excitation is described by the spin operator $\hat{\sigma} _{gs}$ which 
can be found in terms of Eqs.(\ref{eq:as},\ref{eq:ai}):
\begin{align}
& \hat{\sigma}_{gs}(D,\omega) = \frac{\Omega ^* g}{V(\omega)} \left(\hat{a}_S (D,\om)- \frac{\Op}{\Omega ^*} \frac{\omega+i\gge}{\Delta}\hat{a}_I^\dagger(D,\omega) \right). \label{eq:spin}
\end{align}
It has contributions from both the signal and idler field, but the idler term is proportional to the small parameter $\gge/\Delta$, and therefore can largely be ignored compared to the signal part. 
We can therefore concentrate on solving for the propagation of the signal field and our results will still be applicable to EIT quantum memory.  

\begin{figure}[t]
\includegraphics[width=7cm]{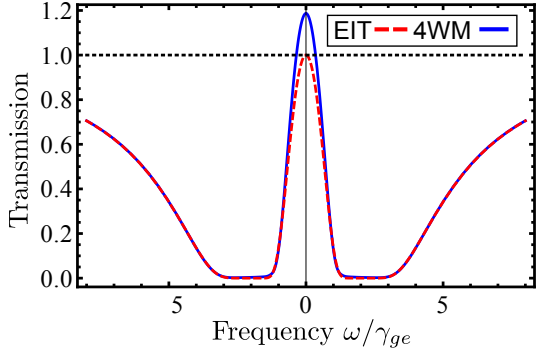}
\caption{
Plot of the transmission coefficient as a function of frequency for EIT with 4WM (solid) and without 4WM (dashed). The parameters were chosen to emphasize the spectral behavior.} 
\label{fig:transmission}
\end{figure}

\begin{figure}[t]
\includegraphics[width=7cm]{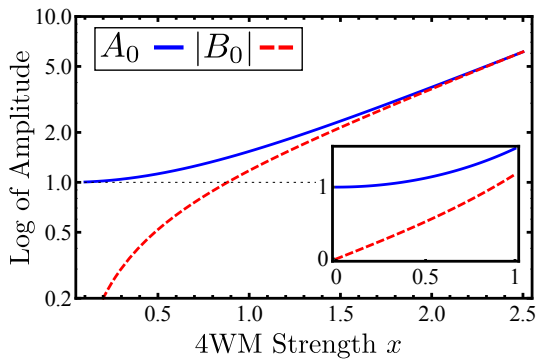}
\caption{
Log plot of the signal (solid) and idler (dashed) amplitudes for parameters $\Omega=0.1\gge$, $\De/\gge=33$ and $\eta=1$ as a function of effective 4WM optical depth.
Assuming no initial idler field and normalized to the amplitude of the initial signal field.
Note both become equal and grow exponentially for high optical depths.
The inset is a linear plot showing the the low optical depth behavior of the amplitudes.
 } 
\label{fig:A0B0}
\end{figure}

To get some intuition for what is going on in the system, we first consider the semi-classical solution for the signal and idler fields.
Since we are mainly interested in quantum memory applications and it has already been shown that an input idler field is not stored \cite{Novikova_2011},
we will assume no input idler field, implying we can ignore $C(D, \om)$. 
In this case, the semi-classical solution for the fields is just: 
\begin{eqnarray}
\alpha_S (D, \omega) &=& A(D, \omega)\,  \alpha_S(0,\omega), \label{eq:alphaS}\\
\alpha_I (D, \omega) &=& -B^{*}(D, \omega) \alpha_S^*(0,\omega). \label{eq:alphaI}
\end{eqnarray}
While the expressions for A and B are complicated, in the limit of $\epsilon \ll 1$ which we have already assumed in order to derive the reduced Hamiltonian and $D > 1$ as is necessary for QM, 
both $A$ and $B$ are well approximated by Gaussians for frequencies near resonance:
\begin{align}
& A(D, \omega) = A_0(D) e^{i\tau_S(D)\omega-\frac{\omega^2}{\Delta\omega_S^2}}, \label{eq:AGauss}\\
& B(D, \omega) = B_0(D) e^{i\tau_I(D)\omega-\frac{\omega^2}{\Delta\omega_I^2}}. \label{eq:BGauss}
\end{align}
where the $\tau_{S}$, $\tau_I$ are the group delay times for the signal and generated idler field, respectively; while $\Delta \omega_S$ and $\Delta \omega_I$ are the frequency widths.   
It is clear that amplitudes $A_0$ and $B_0$ are the steady state solutions for the fields:
\begin{align}
A_0 = & \cosh\biggl(\frac{D\gge \eta}{\Delta}\biggr)\, e^{\frac{D\gge^2 \eta^2}{2\Delta^2}} \nonumber \\
& -\frac{i}{2}\frac{\eta\gge}{\Delta}\sinh\biggl(\frac{D\gge \eta}{\Delta}\biggr)\, e^{\frac{D\gge^2 \eta^2}{2\Delta^2}}, \label{eq:A0} \\
B_0 = & \frac{i\Ops \Omega^*}{|\Op \Omega|}\sinh\biggl(\frac{D \gge \eta}{\Delta}\biggr)\, e^{\frac{D\gge^2 \eta^2}{2\Delta^2}}. \label{eq:B0}
\end{align}
We introduce the ratio of the control field Rabi frequencies $\eta = |\Op|/|\Omega|$, which only differs from unity when the two transitions have different dipole moments.
It is convenient to introduce a parameter to keep track of the effective 4WM strength, 
\begin{align}
x = D \eta \frac{\gge}{\Delta}.
\end{align}
Then for field propagation, 
Eqs.(\ref{eq:A0},\ref{eq:B0}) define two distinct regimes of $x$.
For large 4WM strength $x > 1$, both the signal and idler field experience exponential growth and except for a phase factor are essentially the same, as shown in Fig. \ref{fig:A0B0}.
\begin{align}
& A_0 = \frac{1}{2}e^{x}, \\
& B_0 = \frac{i}{2}\frac{\Ops\Omega^*}{|\Op \Omega|}e^{x}.
\end{align}
At large $x$ the 4WM process is generating many more photons than are in the initial pulse, and for every new signal field photon there is a corresponding idler photon generated. 
While for $x \ll 1$, there is only weak gain for the idler and signal field, i.e. we can treat this perturbatively:
\begin{align}
& A_0 = 1+\frac{x^2}{2}, \\
& B_0 = i\frac{\Ops \Omega^*}{|\Op \Omega|}x. \label{eq:B0small}
\end{align}
The idler field grows faster than the signal, but since it starts from vacuum, it remains much weaker than the signal field.

Since 4WM introduces gain on the signal field, which also leads to stronger matter excitations, when present it will always increase the classical storage and retrieval efficiencies for
an EIT memory. Therefore, in experiments that see a loss of classical efficiencies at higher optical depths such as \cite{Novikova_2008}, the loss should not be attributed to 4WM, but rather
to other processes that grow with optical depth such as increased dephasing or depletion of the control field.   
The case of particular interest for QM will be for small 4WM strengths $x<1$, since we will show in Sect. \ref{sect:noise} that exponential growth of the signal field is accompanied by an equally strong growth in noise. 
%

The group delay time for the fields is given by $\tau_S$ and $\tau_I$, where a field traveling at the speed of light is taken not to have a time delay. In the low optical depth case:
\begin{align}
& \tau_S \simeq \frac{D\gge}{|\Omega|^2}, \\
& \tau_I \simeq \frac{D\gge}{2|\Omega|^2},
\end{align}
such that $\tau_S$ is essentially the standard EIT delay time. As a consequence, the spatial pulse compression with 4WM is the same as standard EIT. The delay time for the generated idler $\tau_I$, is approximately half of that for the signal field.
This can be understood as a consequence of the idler field being generated from the signal field. While the idler field is essentially moving at the speed of light and therefore is not delayed,
it is also constantly being generated by the slow signal field. The total delay time is then the average of delay for idler photons generated near the beginning of the medium, and the idler photons
generated at the end of the medium after the slower signal field has traversed the medium length.  
A similar effect is seen at high optical depth: 
\begin{align}
& \tau_S \simeq \tau_I \simeq  \frac{D\gge}{2|\Omega|^2},
\end{align}
where there is now a locking of the velocities for both fields to the average. We note that $\tau_S$ is now a factor of two smaller than in the ideal EIT case and the pulse compression is even reduced. Both fields are growing exponentially and travel together, the faster idler field is generating a new slower signal field which 
leads to less signal delay.
\begin{figure}
\includegraphics[width=8.5cm]{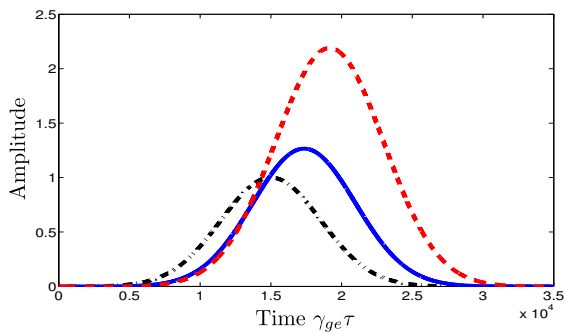}
\caption{
Plot of the signal field at the beginning of the medium (dot-dashed), after an effective optical depth of $x= .75$ (solid), 
and after $x= 3$ (dashed). Normalized to the initial pulse amplitude. The parameters are $\Omega=0.1\gge$, $\De/\gge=33$ and $\eta=1$} 
\label{fig:propagation}
\end{figure}
%
\begin{figure}[t]
\includegraphics[width=7cm]{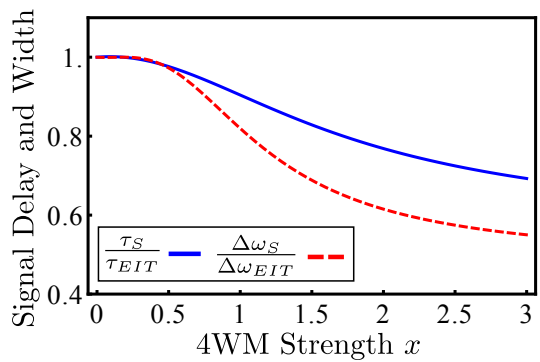}
\caption{
Plot of the signal delay time, normalized to the standard EIT delay time $D\gge/|\Omega|^2$ (solid) and the signal transmission frequency width, 
normalized to the standard EIT transmission window $|\Omega|^2/(\gge \sqrt{D})$ (dashed), as a function of the effective 4WM optical depth.
With $\Omega=0.1\gge$, $\De/\gge=33$ and $\eta=1$. 
 } 
\label{fig:ratios}
\end{figure}

The spectral behavior of the fields is described by the frequency widths $\Delta \omega_S$ and $\Delta \omega _I$, and we can again
distinguish between two different regimes. 
For low optical depth, 
\begin{align}
& \Delta \omega _S \simeq \frac{|\Omega|^2}{\gge \sqrt{D}}, \\
& \Delta \omega _I \simeq  \frac{|\Omega|^2}{\gge \sqrt{D}} \sqrt{\frac{2}{1+D/12}}. \label{eq:B2small}
\end{align}
Where $\Delta \omega _S$ defines the usual EIT transmission window, and $\Delta \omega _I$ defines the transparency window of the idler field.
At high optical depth the widths are given by:
\begin{align}
& \Delta \omega _S \simeq \Delta \omega _I \simeq \frac{|\Omega|^2}{\gge \sqrt{D}} \sqrt{\frac{8\gge \eta}{\Delta}}.
\end{align}
The fields are now propagating together with a similar transmission window that is narrower than the original EIT transmission window by a factor of $\sqrt{8\gge \eta/\Delta}$. 
This narrowing is due to preferential gain of the signal near resonance where it does not experience absorption, rather than at frequencies near the edge of the EIT transmission window that see some absorption. 

We have shown that for low 4WM strengths $x<1$, the signal field propagates similar to normal EIT, with a small gain due to 4WM from the newly generated idler field which remains weak compared to the signal field 
and propagates through the medium as if it was transparent. While, the propagation is dramatically different at high 4WM strengths $x > 1$, 
where exponential growth of the signal and idler field lock the fields together 
such that they have equal amplitudes, experience less group delay, and have a narrower transmission window.  

\section{Signal Intensity and additive photon noise}
\label{sect:noise}

\begin{figure}[t]
\includegraphics[width=7cm]{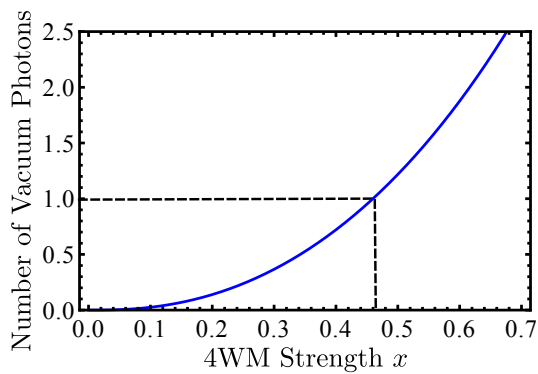}
\caption{
Plot of the number of noise photons produced as a function of effective 4WM optical depth for $\Omega=0.1\gge$, $\De/\gge=33$ and $\eta=1$. Notice it becomes larger than 1 slightly before when $x > 1/2$.
 }
\label{fig:noisephotons}
\end{figure}

As an indicative measure for the effect of 4WM on light storage in an EIT medium 
we now consider the number of signal photons at the end of the medium at some time $\tau$ using Eqs.(\ref{eq:as},\ref{eq:ai}):
\begin{widetext}
\begin{align}
& \expv{\hat{a}_S^\dagger(D,\tau)\hat{a}_S(D,\tau)}=\int\!\!\!\int\!\!{\rm d}\om^\prime 
{\rm d}\om\, e^{-i(\om^\prime-\om)\tau}\biggl[ A^*(D,\om^\prime)A(D,\om)\expv{\hat{a}_S^\dagger(0,\om^\prime)\hat{a}_S(0,\om)} \\ \nonumber
&\quad \quad \quad \quad \quad \quad \quad \quad \quad + B^*(D,\om^\prime)B(D,\om)\expv{\hat{a}_I(0,\om^\prime)\hat{a}_I^\dagger(0,\om)} \biggr]\\
&=\int\!\!\!\int \!\! {\rm d}\om^\prime {\rm d}\om e^{-i(\om^\prime-\om)\tau} \, A^*(D,\om^\prime)A(D,\om)\expv{\hat{a}_S^\dagger(0,\om^\prime)\hat{a}_S(0,\om)} + \int d\om |B(D,\om)|^2,
\end{align}
\end{widetext}
where the first part corresponds to the semi-classical solution, which one would obtain by treating the fields classically with no 
input idler field. The second term contains anti-normally ordered products of the field operators, so we used the commutator 
relation for the field operators $[ \hat{a}_I (\omega^\prime), \hat{a}_I ^\dagger (\omega)] = \delta (\omega - \omega^\prime) $ to bring it back to normal order, 
i.e. this part is a pure quantum mechanical effect. Since the value of the second part is equal for all time 
$\tau$ it describes the generation rate of the incoherent signal photons.
This contribution exists even when there is no signal input at all, and therefore is important for few photon input fields 
consequently, we refer to this as the vacuum noise contribution.
It does not grow with signal field strength, so is much less important for fields with large photon number. 
With this we are able to estimate the number of noise photons by multiplying the generation rate 
with the propagation time of the signal field.  
As noted in Sect. \ref{sect:sol}, at high optical depth $x > 1$, $A_0$ and $B_0$ are equal, which implies that for a single photon input, the vacuum noise contribution will be as strong as the output
of the signal field. Therefore in the regime of $x>1$, a quantum memory is impossible. 
And as one can see from Fig. \ref{fig:noisephotons} there will already be an additional noise photon generated for 4WM strengths near $x=0.5$.

\bigskip
\section{Noise due to finite excited state population}
\label{sect:excited}

Additionally to the vacuum noise, there will also be noise contributions due to spontaneous emission. 
In the following we will calculate these noise contributions, directly from the Langevin noise operators which are second order in the
signal and idler fields.
The noise operators introduced in Eqs.(\ref{eq:as},\ref{eq:ai}) read explicitly:
\begin{align}
&  \delta \hat \alpha _S(\xi, \omega) = \int _0 ^{\xi} \! \! \! \!  d\xi ^\prime A(\xi - \xi ^\prime, \omega) \hat{F} _S + \int _0 ^\xi \! \! \! \!  d\xi^\prime B(\xi-\xi^\prime, \omega) \hat{F} _I, \\
&  \delta \hat \alpha _I(\xi, \omega) = - \! \int _0 ^{\xi} \! \! \! \!  d\xi ^\prime B(\xi - \xi ^\prime, \omega) \hat{F} _S + \int _0 ^\xi \! \! \! \!  d\xi^\prime C(\xi-\xi^\prime, \omega) \hat{F} _I,
\end{align}
where $A$, $B$, and $C$ are given by Eqs.(\ref{eq:fullA}-\ref{eq:fullC}), and the new $\hat{F}$ operators are defined in terms of the atomic Langevin noise operators:
\begin{align}
\hat{F}_S =& \frac{gN \Omega}{(\omega + i\gamma_{gs}) (\omega +i\gamma _{ge})-|\Omega|^2} \hat{F}_{gs} \nonumber \\
&-\frac{g N (\omega+i\gamma _{gs})}{(\omega + i\gamma_{gs}) (\omega +i\gamma _{ge})-|\Omega|^2} \hat{F}_{ge} \\
\hat{F}_I =& \frac{gN \Omega ^{\prime *} (\omega+i\gge)/\Delta}{(\omega + i\gamma_{gs}) (\omega +i\gamma _{ge})-|\Omega|^2} \hat{F}_{gs} \nonumber \\
&-\frac{g N \Omega^* \Omega ^{\prime *}/\Delta}{(\omega + i\gamma_{gs}) (\omega +i\gamma _{ge})-|\Omega|^2} \hat{F}_{ge}
\end{align}
The rate of generation for extra amplitude noise due to spontaneous emission, is given by the expectation value $\langle \delta \hat \alpha _S ^\dag \delta \hat \alpha _S \rangle$ for which to the first non-zero order in $\Omega ^\prime/\Delta$ is:
\begin{widetext}
\begin{align}
\langle \delta \hat \alpha _S ^\dag \delta \hat \alpha _S \rangle = \frac{1}{2\pi} \int _{-\infty} ^{\infty} \! \! \! \! \! \! d\omega \int _{-\infty} ^{\infty} \! \! \! \! \! \! d\omega^\prime  
\int _0 ^D \! \! \! \! d\xi \int _0 ^D \! \! \! \! d\xi ^\prime A^*(D-\xi,\omega) A(D-\xi ^\prime,\omega^\prime)  \langle \hat{F}_S ^\dag(\xi,\omega) \hat{F}_S(\xi ^\prime,\omega^\prime) \rangle e^{i(\omega - \omega ^\prime)t} . \label{eq:asas}
\end{align}
While the form of the $\hat{F}$ are unknown, their correlations can be calculated using the fluctuation dissipation theorem \cite{Lamb_1974}, but we need to consider the full set of equations from the
reduced Hamiltonian given in Eq.(\ref{eq:H}), not just the $\sge$ and $\sgs$ equations Eqs.({\ref{eq:sge1},\ref{eq:sgs1}}). The full equations, neglecting detunings, letting $\ggs = 0$,
and dropping terms that are third power in the fields are:
\begin{align}
& \dot{\hat{\sigma}} _{ge} =  i\omega \hat{\sigma}_{ge}-\gge\hat{\sigma}_{ge}+ig\hat{a}_S +i\Omega \hat{\sigma}_{gs} + \hat{F}_{ge},\\ \label{eq:ge2}
& \dot{\hat{\sigma}} _{se} =  -\gse\hat{\sigma}_{se}+i\Omega(\hat{\sigma}_{ss}-\hat{\sigma}_{ee}) -i\frac{\Ops g\hat{a}_I}{\D}\hat{\sigma}_{ge} +ig\hat{a}_S \hat{\sigma }_{gs}^\dag + \hat{F}_{se}, \\
& \dot{\hat{\sigma}} _{gs} =   i\omega \hat{\sigma}_{gs} + i\frac{\Op g \hat{a}_I^*}{\D} +i\Omega^* \hat{\sigma}_{ge} + \hat{F}_{gs},\\
& \dot{\hat{\sigma}} _{ee} = -r_{es}\hat{\sigma}_{ee} -r_{eg}\hat{\sigma}_{ee} -ig\hat{a}_S^{\dag} \hat{\sigma}_{ge} +ig\hat{a}_S\hat{\sigma}_{ge}^{\dag} -i\Omega^* \hat{\sigma}_{se} +i\Omega \hat{\sigma}_{se}^{\dag} +\hat{F}_{ee}, \\
& \dot{\hat{\sigma}} _{ss} = r_{es} \hat{\sigma}_{ee} -r_{sg}\hat{\sigma}_{ss} -i\frac{\Ops}{\D} g \hat{a}_I^{\dag *} \hat{\sigma}_{gs} +i\frac{\Op }{\Delta}g \hat{a}_I^*\hat{\sigma}_{gs}^{\dag} +i\Omega^* \hat{\sigma}_{se} -i\Omega \hat{\sigma}_{se}^{\dag} +\hat{F}_{ss}. \label{eq:ee2}
\end{align}
\end{widetext}

\begin{figure}[t]
\includegraphics[width=7cm]{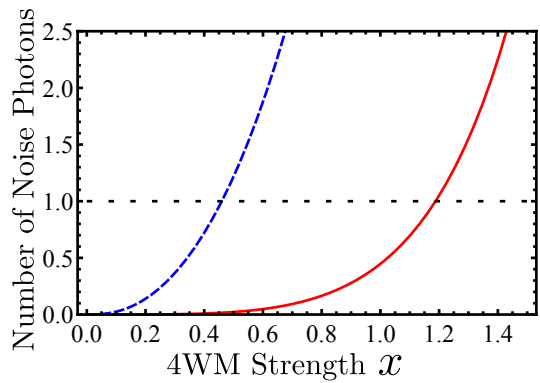}
\caption{
Number of photons affected by spontaneous emission (solid) showing at what effective 4WM optical depth it becomes larger than 1, assuming a single photon input. 
For comparison the number of vacuum noise photons (dashed) is also plotted. In both cases the parameters are $\Omega=0.1\gge$, $\De/\gge=33$ and $\eta=1$}
\label{fig:comparison}
\end{figure}

Notice that the equations are no longer linear, but we expect the noise terms to be second order in the fields.  
From Eqs.(\ref{eq:ge2}-\ref{eq:ee2}) we can find the correlations of the Langevin noise operators:
\begin{align}
& \dude{ \hat{F}_{ge} ^\dag \hat{F}_{ge}} = (2\gge - r_{es} -r_{eg})\dude{\hat{\sigma}_{ee}}, \\
& \dude{ \hat{F}_{gs} ^\dag \hat{F}_{gs}} = r_{es}\dude{\hat{\sigma}_{ee}}, \\
& \dude{ \hat{F}_{ge} ^\dag \hat{F}_{gs}} = (\gge  - \gse)\dude{\hat{\sigma}_{se}^{\dag}}, \\
& \dude{ \hat{F}_{gs} ^\dag \hat{F}_{ge}} = (\gge  - \gse)\dude{\hat{\sigma}_{se}}.
\end{align}
With these values we can calculate $\dude{\hat{F}_S^{\dag} \hat{F}_S}$:
\begin{align}
& \dude{\hat{F}_S^{\dag}(\xi,\omega) \hat{F}_S(\xi^\prime,\omega^\prime)} = \frac{g^2 N^2 \delta (\omega - \omega ^\prime) \delta(\xi-\xi^\prime)}{|\omega(\omega-i\gge)-|\Omega|^2|^2}\biggl[ \nonumber \\
& r_{es}|\Omega|^2\dude{\hat{\sigma}_{ee}} + (\gse-\gge) \dude{\hat{\sigma}_{se}} \omega \Omega ^* +\nonumber \\
&(\gse-\gge) \dude{\hat{\sigma}_{se}^{\dag}} \omega \Omega +\omega^2 (2\gge -r_{es}-r_{eg})\dude{\hat{\sigma}_{ee}} ) \biggr]. \label{eq:FsdFs}
\end{align}
In Eq.(\ref{eq:FsdFs}) we can neglect
the contributions from $\dude{\hat{\sigma}_{se}}$ under the reasonable assumption that $|\gse - \gge| \ll \gse$, i.e.
only the excited state population is important.
Then the average excited state population can be found by solving the semi-classical form of Eqs.(\ref{eq:ge2}-\ref{eq:ee2}):
\begin{align}
\dude{\hat{\sigma}_{ee}} = & \frac{2g^2\gge \omega^2|\alpha_S|^2}{r_{eg} |V(\omega)|^2} - \frac{2g^2\gge \omega(|\Op \Omega|/\Delta)\sqrt{|\alpha_S \alpha_I|}}{r_{eg} |V(\omega)|^2} \nonumber \\
& +  \frac{2g^2\gge |\Omega|^2 (|\Op|^2/\Delta^2) |\alpha_I|^2  }{ r_{eg} |V(\omega)|^2},
\end{align}
where $\alpha _S$ and $\alpha _I$ are the semi-classical field solutions given by Eqs.(\ref{eq:alphaS},\ref{eq:alphaI}).
Assuming the initial field is given by a Gaussian distribution: 
\begin{align}
|f(\omega)|^2 = \frac{1}{\sqrt{\pi} \Delta \omega _0} e^{-(\frac{\omega}{\Delta \omega _0})^2}, \label{eq:F}
\end{align}
where $\Delta \omega _0$ is the frequency width of the incoming pulse; 
and then multiplying by the delay time of the signal field $\tau_D$, yields the number of noise photons affected by dephasing due to spontaneous emission:
\begin{widetext}
\begin{align}
& \mathcal{N}_{\rm{SE}} = \tau_D \dude{\delta \hat{\alpha}_S ^{\dagger} \delta \hat{\alpha}_S } = \tau_D \int _{-\infty} ^{\infty} \! \!\! \! \! {\rm d}\omega \!\int _0 ^{D}\!\! {\rm d}\xi |A(D-\xi,\omega)|^2 |f(\omega)|^2 \frac{g^4N^2[(2\gge-r_{es}-r_{eg})\omega^2 +r_{es}|\Omega|^2]}{r_{eg}[\gge ^2 \omega^2+|\Omega|^4]^2} \nonumber \\
& \times\biggl[ 2\gge \omega^2 |A(\xi, \omega)|^2 -2\gge \omega \frac{|\Op \Omega|}{\Delta}|A(\xi,\omega)||B(\xi,\omega)| + 2\gge |\Omega|^2 \frac{|\Op|^2}{\Delta ^2} |B(\xi,\omega)|^2 \biggr].
\end{align}
The integral over $\xi$ is straight forward when we use the Gaussian approximation developed in Sect. \ref{sect:sol} for $A(\xi,\omega)$ and $B(\xi,\omega)$.
Approximating $|A_0(\xi)| = \cosh (\gge \eta \xi /\Delta)$ and $|B_0(\xi)| = \sinh (\gge \eta \xi /\Delta)$, and performing the $\xi$ integral leaves us with:
\begin{align}
& \mathcal{N}_{\text{SE}} = \tau_D \int _{-\infty} ^{\infty} \! \! \! \! \! \! d\omega e^{-2(\omega/\Delta \omega_S)^2} |f(\omega)|^2 \frac{g^2N^2[(2\gge-r_{es}-r_{eg})\omega^2 +r_{es}|\Omega|^2]}{r_{eg}[\gge ^2 \omega^2+|\Omega|^4]^2}  \nonumber \\
& \times\biggl[ 2\gge \omega^2 e^{-2(\omega/\Delta \omega_S)^2} \Bigl(-\frac{D}{4} + \frac{D}{8}\cosh(x) + \frac{5D}{16x}\sinh(2x)\Bigr)  \nonumber \\
& \quad -2\gge \omega \frac{|\Op \Omega|}{\Delta}e^{-(\omega/\Delta \omega_S)^2}e^{-(\omega/\Delta \omega_I)^2} \sinh(x)\Bigl(\frac{D}{4}\cosh(x)+\frac{1}{4}\sinh(x)\Bigr) \nonumber \\
& \quad + 2\gge |\Omega|^2 \frac{|\Op|^2}{\Delta ^2} e^{-2(\omega/\Delta \omega_I)^2} \Bigl(-\frac{D}{4} + \frac{D}{8}\cosh(2x)+ \frac{D}{16x}\sinh(2x)\Bigr) \biggr]
\end{align}
Now consider that the spectral window of $A(D,\omega)$ will ensure that the main contribution of the integral comes for small $\omega$, 
so we can take $\omega < |\Omega|^2/\gge$.  
This will allow us to approximate by replacing $\gge ^2 \omega^2+|\Omega|^4$ with $|\Omega|^4$ and perform the $\omega$ integral:
\begin{align}
& \mathcal{N}_{\text{SE}} \simeq \frac{g^4 N^2r_{es} \tau_D}{r_{eg}|\Omega|^6} \frac{D}{4}\, \biggl[ \Delta \omega _0 ^2 \gge \Bigl(\frac{1}{2}\cosh(x)-1
+ \frac{5}{4x}\sinh(2x)\Bigr)
+ \gge |\Omega|^2 \frac{|\Op|^2}{\Delta ^2}  \Bigl(\cosh(2x)-2 + \frac{1}{2x}\sinh(2x) \Bigr) \Bigr]
\end{align}
\end{widetext}
This can be simplified by taking $r_{es} = r_{eg}$, and by noticing that the square of the group index $n_g^2 = g^4N^2/|\Omega|^2$ is approximately equal
to $1/(\gge \tau_D)$. We can now estimate the number of dephased photons due to spontaneous emission in our two limits,
first in the small optical depth regime, where $x < 1$, we can simplify further:
\begin{align}
\mathcal{N}_{\text{SE}} \simeq  \frac{D}{2} \biggl[ \frac{\Delta \omega _0 ^2}{|\Omega|^2} (1+\frac{x^2}{8})
+  \frac{|\Op|^2}{\Delta ^2}  (\frac{1}{2} + x^2) \biggr].
\end{align}
Then for large optical depth, $x \gg 1$, we have:
\begin{align}
& \mathcal{N}_{\text{SE}} \simeq \frac{D}{16}  e^{2x} \biggl[ \frac{\Delta \omega _0 ^2}{|\Omega|^2} \frac{5}{2x} +  2\frac{|\Op|^2}{\Delta ^2}  
\biggr],
\end{align}
which like the vacuum noise is exponentially growing. Note that the noise induced by spontaneous emission is proportional to the number of photons in the initial signal field, thus
for a small number of initial photons its contribution will be much weaker than the vacuum noise contribution derived in Sect. \ref{sect:noise}, as illustrated in Fig. \ref{fig:comparison}. 
Therefore, we can neglect this effect in the single photon fidelity calculation of
Sect. \ref{sect:results}. Of course, for a classical field with a large number of photons, dephasing due to spontaneous emission will be the dominant contribution to the noise. 

\section{Fidelity of Propagation}
\label{sect:results}

The figure of merit for a QM is the fidelity. For a wave propagating through the medium, we can calculate the 
fidelity for a particular input field in a pure state $\ket{\Psi_{\text{in}}}$,
by finding the overlap of the wave-function well before the medium and the wave-function well after the medium.
The fidelity is then defined as the infimum of the square root of the overlap over the set of all possible input functions.
\begin{align}
& F_{\ket{\Psi_{\text{in}}}} = \inf\limits_{\ket{\Psi_{\text{in}}}} \sqrt{\bra{\Psi_{\text{in}}} \rho _{\text{out}} \ket{\Psi_{\text{in}}}}.
\label{eq:fid}
\end{align}
To be able to calculate this overlap we need to know the state of the output field described by $\hat{\rho}_{\text{out}}$. To extract the output state from our field operator solutions given by 
Eqs. (\ref{eq:as}, \ref{eq:ai}) we first write the density matrix in the Glauber P-representation. Then using the operator solutions we calculate the normally ordered characteristic function
which is in turn the Fourier transform of the Glauber P-function. The operator solutions are given in the frequency domain and to facilitate the calculation we discretize the frequency space
into $2M+1$ modes centered at the resonance frequencies.

We know that outside of the medium the fields obey free evolution, therefore we can separate space into 3 different regions: the region before the medium, the region after the medium and the 
region inside the medium. Then we can write for the signal field in the first region:
\begin{align}
 \hat{a}_S^{\rm in}(\tau) = \frac{\De\omega}{2\pi}\sum_n \hat{c}_{{\rm in},n} e^{-i\tau \omega_n}, 
\end{align}
where $\De\omega=\frac{c}{L_Q}\frac{1}{2M+1}$ is the frequency spacing for some quantization length $L_Q$ and $\hat{c}_{{\rm in},n}$ is an annihilation operator, 
which destroys a photon with frequency $\omega_n= \omega_S+n\De \omega$ in the region before the medium. For the signal field in the region after the medium we have a similar expression:
\begin{align}
 \hat{a}_S^{\rm{out}}(\tau) = \frac{\De\omega}{2\pi}\sum_n \hat{c}_{{\rm out},n} e^{-i\tau\omega_n},
\end{align}
where $\hat{c}_{{\rm out},n}$ is now the operator which annihilates the photon with frequency $\omega_n= \omega_S + n \De \omega$ in the region after the medium. The same argumentation holds 
for the idler field with corresponding idler operators $\hat{b}_{\rm{in},n}$ and $\hat{b}_{\rm{out},n}$. Now  the mapping between the \textit{in} and \textit{out}
operators is given by:
\begin{align}
& \hat{c} _{{\rm out},n} = A_n \hat{c}_{{\rm in},n} + B_n \hat{b}^{\dag}_{{\rm in},n}, \label{eq_map}
\end{align}
where $A_n = A(\omega _n, D)/\sqrt{2M+1}$ and $B_n = B(\omega_n, D)/\sqrt{2M+1}$.

We are primarily interested in the fidelity for an input state that is a superposition of states containing either zero photons or a single photon: 
\begin{align}
& \ket{\Psi_{\text{SP}}} = C_0 \ket{\{ 0 \}} + C_1 |\{1\}\rangle \nonumber \\
& =C_0 \ket{\{ 0 \}_S} \ket{\{ 0 \}_I} + C_1\sum _{n} f_n\, {\hat c}^{\dag} _n \ket{\{ 0 \}_S} \ket{\{ 0 \}_I}, \label{eq:fullstate}
\end{align}
where $C_0$ and $C_1$ are constants with $|C_0|^2+|C_1|^2=1$,  
$f_n$ is a distribution function that represents the photons frequency envelope, and is normalized to produce a single photon by:
\begin{align}
\sum _{n} |f_n|^2 = 1.
\end{align}
While $\ket{\{ 0 \}_S} = \prod_n \ket{0_n}_S$, $\ket{\{ 0 \}_I} = \prod_n \ket{0_n}_I$ are the vacuum product states for the signal and idler fields.
Our calculations show that the fidelity is always lower when the incoming state is purely a single photon, 
due to any noise that is detrimental to the vacuum input
being equally detrimental to the single photon state. 
Therefore, the infimum is attained for $C_1 = 1$ and $C_0 = 0$, which we will use in all further
calculations. 

In order to calculate the fidelity we will first write the overlap of the single photon input from Eq.(\ref{eq:fid}) in terms of
the multimode Glauber P-representation $P(\beta_n)$, i.e. basically using an expansion of the density matrix in coherent states $\ket{\beta_n}$: 
\begin{align}
\bra{\Psi_{\text{SP}}} \hat{\rho} _{\text{out}} \ket{\Psi_{\text{SP}}} = \prod _{n} \int _{-\infty} ^{\infty} \! \! \! \! \! \! \! d^2 \beta _n  
|\braket{\{ 1\}}{\{ \beta \}}|^2 P(\beta_n).
\end{align}
Here $|\braket{\{ 1\}}{\{ \beta \}}|^2$ is the overlap of the single photon state with the multimode coherent state, 
which can be found by expanding the coherent states as an infinite sum of Fock states:
\begin{align}
& |\braket{\{ 1\}}{\{ \beta \}}|^2 = (\prod_n e^{-|\beta_n|^2}) (\sum _{j,k} f_j^* f_k \beta_j \beta_k^*).
\end{align}
The Glauber P-function in turn is the inverse Fourier-transform of the normally ordered characteristic function:
\begin{align}
P(\beta_n) = \! \prod _{n} \frac{1}{\pi^2} \int _{-\infty} ^{\infty} \! \! \! \! \! \! \! d^2 \phi_n 
e^{\phi_n^* \beta_n -\phi_n \beta_n^*} \chi_N,
\end{align}
The multimode characteristic function \cite{Barnett_2002} can be found from the trace over the density matrix using the operator input-output relations given in Eq.(\ref{eq_map}):
\begin{widetext}
\begin{align}
& \chi _N = \bra{\Psi_{\text{SP}}} \exp \left(+\sum_{n} \phi_n \hat{c}_{out,n} ^{\dag} \right)
\exp \left( -\sum _{m} \phi_m ^* \hat{c}_{out,m}  \right) \ket{\Psi_{\text{SP}}}. \label{eq:chiN}
\end{align}
\end{widetext}
Putting it all together reduces the fidelity calculation to finding the normally ordered characteristic function and performing integrals over it:
\begin{align}
& \left(F _{\text{SP}}\right)^2 =  \prod _{n} \frac{1}{\pi^2} \int _{-\infty} ^{\infty} \! \! \! \! \! \! d^2 \beta _n
\int _{-\infty} ^{\infty} \! \! \! \! \! \! \! d^2 \phi_n 
|\braket{\{ 1\}}{\{ \beta \}}|^2 e^{\phi_n^* \beta_n -\phi_n \beta_n^*} \chi_N \label{eq:integral}
\end{align}
Using Eq.(\ref{eq_map}) in Eq.(\ref{eq:chiN}) and taking the expectation value over the single photon state given 
by Eq.(\ref{eq:fullstate}) we can calculate the characteristic function. The problem nicely breaks up into 
finding the expectation of the operators associated with the signal field
and the expectation of the operators associated with the idler field, thus we can take:
\begin{equation}
\chi _N = \chi _N ^{\text{sig}} \chi _N ^{\text{vac}},
\end{equation}
where we find that:
\begin{align}
& \chi _N ^{\text{sig}} = 1 - \sum _{n,m} f_n^* f_m  \phi_n A_n^* \phi_m^* A_m, \\
& \chi _N ^{\text{vac}} = \exp \left( -\sum_{n} |\phi_n|^2 |B_n|^2 \right).
\end{align}
Performing the integrals of Eq.(\ref{eq:integral}) while being careful with the sums gives the single photon fidelity as:
\begin{align}
& \left(F_{\text{SP}}\right)^2 = \left( \prod _{n=-M} ^M \frac{1}{1+|B_n|^2} \right) \Biggl[ \sum _{i,j} \frac{|f_i|^2 |f_j|^2 A_i^* A_j}{(1+|B_i|^2)(1+|B_j|^2)} \nonumber \\
&+ \sum _{i} \frac{|f_i|^2  |B_i|^2}{(1+|B_i|^2)} 
- \sum _{i,j} \frac{ |f_i|^2 |f_j|^2 |A_i|^2 |B_j|^2}{( 1+|B_i|^2)(1+|B_j|^2 )} \Biggr] \label{eq:fidelity}
\end{align}
Eq.(\ref{eq:fidelity}) has two parts, a product multiplied by a sum. 
The product can be interpreted as the vacuum contribution, it is the same as would be calculated for a vacuum state.
It always converges and depends on the spectral width of the coupling coefficient. In the continuous limit it can be explicitly calculated:
\begin{align}
\prod _{n=-M} ^M \frac{1}{1+|B_n|^2} \rightarrow \exp \Biggl(-\frac{\tau_S}{2} \int _{-\infty} ^{\infty} \! \! \! \! \! \! d\omega |B(D,\omega)|^2 \Biggr).
\end{align}
For large optical depths $|B_0|^2 \gg 1$, this term dominates the single photon fidelity quickly dropping it to zero, with a rate that is at least exponential in optical depth.
The second part is given by the sums in Eq.(\ref{eq:fidelity}), and is due to the gain on the signal field which is also detrimental for our definition of fidelity if it leads
to having more than a single photon. In the continuum limit $M \rightarrow \infty$ the sums can be converted back into integrals:
\begin{widetext}
\begin{align}
& \sum _{i} \frac{|f_i|^2  A_i^* }{(1+|B_i|^2)} \rightarrow \int _{-\infty} ^{+\infty} d\omega |f(\omega)|^2  \frac{A^*(\omega)}{(1+|B(\omega)|^2)} \nonumber  \\
&\qquad \simeq \frac{A_0^*}{1+|B_0|^2} \sqrt{\frac{1}{1+\frac{\Delta \omega_0^2}{ \Delta \omega _S^2}-2\frac{|B_0|^2}{1+|B_0|^2} \frac{\Delta \omega _0^2}{ \Delta \omega _I ^2}} }
\exp \left( -\frac{\tau_S^2 \Delta \omega _0^2}{4} \frac{1}{1 + \frac{|B_0|^2}{1+|B_0|^2} \frac{\Delta \omega_0^2}{ \Delta \omega _I^2 }} \right), \label{eq:sumA} \\ 
& \sum _{i} \frac{|f_i|^2  |A_i|^2 }{(1+|B_i|^2)} \rightarrow \int _{-\infty} ^{+\infty} d\omega |f(\omega)|^2  \frac{|A(\omega)|^2}{(1+|B(\omega)|^2)}
\simeq \frac{|A_0|^2}{1+|B_0|^2} \sqrt{\frac{1}{1+2\frac{\Delta \omega_0^2}{ \Delta \omega _S^2}-2\frac{|B_0|^2}{1+|B_0|^2} \frac{\Delta \omega_0 ^2}{\Delta \omega _I ^2}}}, \\ 
& \sum _{j} \frac{|f_j|^2  |B_j|^2 }{(1+|B_j|^2)} \rightarrow \int _{-\infty} ^{+\infty} d\omega |f(\omega)|^2  \frac{|B(\omega)|^2}{(1+|B(\omega)|^2)}
 \simeq \frac{|B_0|^2}{1+|B_0|^2} \sqrt{\frac{1}{1 + \frac{1-|B_0|^2}{1+|B_0|^2} \frac{\Delta \omega_0 
^2}{\Delta \omega _I ^2}}}.
\end{align}
\end{widetext}
Where we have assumed that $A(\omega)$ and $B(\omega)$ are Gaussians given by Eqs.(\ref{eq:AGauss}, \ref{eq:BGauss}) 
and that our initial distribution is Gaussian and given by Eq.(\ref{eq:F}).

Now lets consider the ideal EIT regime where the frequency width of the incoming pulse fits well inside the transmission window:
\begin{align}
& \frac{\Delta \omega _0}{\Delta \omega _S} \ll \sqrt{\frac{2}{D}},
\end{align}
where we have chosen this limit such that the exponential term in Eq.(\ref{eq:sumA}) can be dropped. Then consider two different limits for the 4WM strength,
for $x>1$ the fidelity exponentially decreases, while for small $x$ and keeping only the first order terms we can express the fidelity as: 
\begin{align}
F_{\text{SP}} = \exp\biggl(- \sqrt{3} D\frac{|\Omega '|^2}{\Delta^2}\biggr) \sqrt{ 1 - \frac{\Delta \omega _0^2}{\Delta \omega _S^2} - x^2}.
\end{align}
This shows that 4WM will always degrade single photon fidelity, although this is the expected result since without 4WM,
we have standard EIT propagation, which for narrow pulse spectrum has fidelity close to 1.
 
Therefore when 4WM is unavoidable, it would be best to implement quantum memory in the regime $x = D\eta \gge /\Delta < 1$. 
Even for the very high optical depths required by EIT QM, this can be accomplished by choosing the field polarizations such that $\Op$ only couples to transitions that have
a small $\eta$ or a very large $\Delta$. In the case where $\Op$ couples to the signal transition, such that $\eta \approx 1$ and $\Delta$ is fixed,
it is still possible to lower the effects of 4WM by minimizing the optical decoherence rate $\gge$ to decrease the $\gge/\Delta$ ratio. 
For example, experiments in cold Rb gas trapped in a magneto-optical trap where $\gge$ is just half of the spontaneous emission rate have a ratio
of $\Delta/\gge = 500$ for Rb$^{85}$ and $\Delta/\gge = 1000$ for Rb$^{87}$, making it possible to reach very high optical depths while maintaining a small $x$.
This is supported by experiments such as \cite{coldgas} which saw no signs of 4WM even at large optical depth, $D\approx 150$. 4WM does become important for experiments in warm gases,
especially when a buffer gas is used to lower the spin decoherence time, since then at high densities the self broadening and buffer gas broadening due to
collisions can make $\gge$ significantly larger than the spontaneous decay rate, leading to low ratios of $\Delta/\gge \approx 50$. In this case the noise
due to 4WM will have a significant effect on the fidelity, which has likely been observed in warm Rb gas experiments that measure the fidelity
rather then just the storage efficiency such as \cite{Eden1_2009,Eden2_2009}.   
\begin{figure}[t]
\includegraphics[width=7cm]{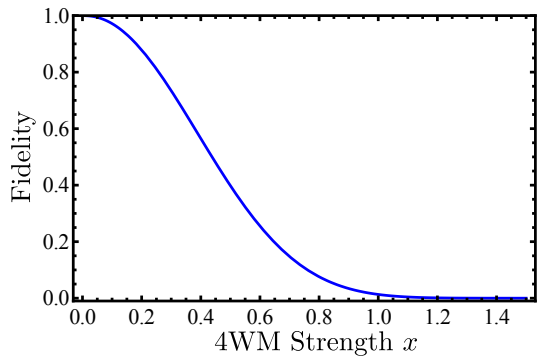}
\caption{
Plot of fidelity for the single photon case as a function of the effective 4WM optical depth for $\Omega=0.1\gge$, $\De/\gge=30$ and $\eta=1$.
In the limit $\Delta \omega_0 /\Delta \omega_S, \Delta \omega_0/\Delta \omega_I \rightarrow 0$.
 }
\label{fig:fidelity}
\end{figure}

\section{Fidelity of Propagation with losses}
\label{sect:losses}

As we have seen in the previous section, the presence of gain due to 4WM leads to a fast reduction of fidelity in otherwise loss-less propagation. We now analyze whether 4WM could be beneficial when there are some linear losses due to scattering in the medium.
In particular, we consider the case where the 4WM gain exactly compensates linear loss, for
which we will compare the EIT and 4WM single photon fidelities.

Our analysis in Sect.\ref{sect:results} also applies to EIT in the limit of $A_0 \rightarrow 1$ and $B_0 \rightarrow 0$. In order to model linear losses in EIT with spatial loss coefficient $\lambda$ we just need to take the expression for the single photon fidelity 
and replace the coefficients with $A_0 = \exp (-\lambda D/2)$ and $B_0=0$.
In that limit the only integral needed is much simpler with:
\begin{align}
&\int _{-\infty} ^{\infty} \! \! \! \! d\omega |f(\omega)|^2 A(\omega) = \nonumber \\  
&A_0 \sqrt{\frac{\Delta \omega _S ^2}{\Delta \omega _0 ^2 + \Delta \omega _S ^2}} \exp \left(-\frac{\Delta \omega _0^2 \Delta \omega_S^2}{4(\Delta \omega_0 ^2+ \Delta \omega _S ^2)}\tau_S ^2 \right),
\end{align}
which in the limit of $\Delta \omega _0 /\Delta \omega _S \ll \sqrt{2/D}$, collapses the fidelity of EIT with losses to the expected result of:
\begin{align}
F_{\text{SP}} ^{\text{EIT}} \simeq |A_0| = e^{-\lambda D/2}.
\end{align}
The same loss can be added to the 4WM fidelity by taking $A_0 \rightarrow A_0 \exp (-\lambda D/2)$ and $B_0 \rightarrow B_0 \exp (-\lambda D/2)$. 

\begin{figure}[t]
\includegraphics[width=7.5cm]{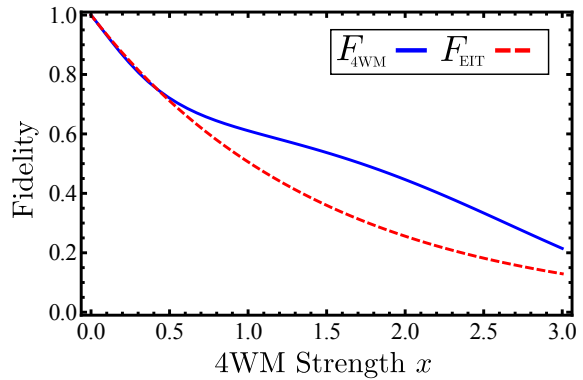}
\caption{
Plot of fidelity of four wave mixing (solid) and standard EIT (dashed) fidelities as a function of effective optical depth in the presence of linear losses,
under the assumption that the signal field is always well within the transmission window. 
The losses are taken to match the 4WM gain present at the effective optical depth of $x=2$. The remaining parameters are $\Delta/\gge = 30$, $\Omega=0.1\gge$ and $\eta=1$. 
} 
\label{fig:withloss}
\end{figure}

We pose the question of when does the 4WM single photon fidelity surpass that of the the EIT, assuming both systems experience linear loss.
In the limit of $\Delta \omega _0 /\Delta \omega _S \rightarrow 0$, we can approximate the fidelity as:
\begin{align}
F_{\text{SP}} = \sqrt{ \frac{ |A_0|^2 (1 - |B_0|^2)}{(1+|B_0|^2)^2} + \frac{|B_0|^2}{1+|B_0|^2}} e^{-\frac{\tau_S}{4} \Delta \omega _I |B_0|^2}  
\end{align}
Now we can take the steady state solution $|A_0| = q \cosh (x)$ and $|B_0| = q \sinh (x)$, with the EIT fidelity given by $q=\exp(-\lambda D/2)$. 
In this case it is fairly simple to calculate when 4WM can improve over EIT, 
it is possible when $q \leq 1/\sqrt{2}$, i.e. the EIT fidelity is already less than 0.7, as illustrated in Fig. \ref{fig:withloss}. 
So while 4WM can be an improvement, it only helps in cases where the fidelity is already too low to use as a quantum memory.

\section{Conclusion}
\label{sect:conclusion}

We developed a model for a field propagating inside an EIT medium that has 4WM. 
We found that there are limits on the use of 4WM EIT as a single photon quantum memory based just on the propagation fidelity. 
We studied the two main sources of noise. The first is due to extra photons generated directly from the vacuum due to 4WM gain.
The second source comes from the finite population that 4WM adds to the excited state, leading to dephasing of the dipoles due to spontaneous emission. 
Together both sources of noise become exponentially large for optical depths $D > \Delta |\Op| /(\gge |\Omega|)$.
This gives a natural limit on how high optical depth can be in EIT based quantum memories when 4WM is present.
In particular, the use of linearly polarized fields in hot gas EIT QM experiments may create difficulties, since the limit can be lower than the
optical depth required for high fidelity QM.

By calculating the fidelity for single photon propagation we can quantitatively describe the degradative effects of 4WM on EIT QM.
We further show that even in the best case scenario where the gain from 4WM compensates some natural losses in the system, for example due to scattering, the propagation fidelity
of 4WM EIT is still worse than that for standard EIT unless the EIT fidelity is below $1/\sqrt{2}$. Therefore for an EIT quantum memory, it is always
preferential to avoid four-wave mixing. This can be accomplished by either choosing field polarizations such that the control field can not 
couple to any nearby transitions, or by working to keep the optical decoherence low to minimize the ratio of $\gge/\Delta$, which is easier to achieve in
low temperature systems.    

Our model so far only considers propagation of the fields through the medium. Since we do not consider the storage process where the control field is turned
off and on, we neglect two considerations. First that it is actually the collective spin excitation that gets stored in an EIT memory, 
which in addition to the signal field contribution contains a small admixture of the idler field as well.
Secondly, we neglect
the limits imposed by needing the field to be wholly within the EIT medium at the time where the control field is turned off for storage, 
i.e. neglecting any field leakage. 
While considering these effects would not improve the limit 4WM imposes on the optical depth, it is possible that when considering them, 
there are scenarios for lower optical depths where 4WM could be made useful. 
We plan to further investigate the effect of 4WM on quantum memory, in particular by finding the effect of 4WM on the collective spin state and considering
the entire storage process. 

\begin{acknowledgments}
The authors acknowledge financial support by the German Federal Ministry of Education and Research (BMBF, project QuOReP 01BQ1005).
\end{acknowledgments}

\bibliographystyle{apsrev4-1}
\bibliography{EIT-4WM_ArXiV}

\end{document}